# Anyons in three dimensions with geometric algebra

Alexander SOIGUINE[1]

[1] SOiGUINE Quantum Computing, Aliso Viejo, CA 92656, USA

**Email address:**
alex@soiguine.com

**Abstract:** Even though it has been almost a century since quantum mechanics planted roots, the field has its share of unresolved problems. Could this be the result of a wrong mathematical structure providing inadequate understanding of the quantum phenomena [1]? Part of the problem is that the terms "state", "observable", "measurement" require a clear unambiguous definition that will make them universally acceptable in both classical and quantum mechanics. This concrete definition will help to further develop a feasible formalism for the challenging area of quantum computing [2].

> ב. לֹא תִהְיֶה אַחֲרֵי רַבִּים לְרָעֹת וְלֹא תַעֲנֶה עַל רִב לִנְטֹת אַחֲרֵי רַבִּים לְהַטֹּת׃
>
> Exodus 23:2

> *"We all say so, and so it must be true"*
> Rudyard Kipling, *The Jungle Book*

## 1. Introduction

A lot of confusion comes from the lack of precision in using terms like "state", "observable", "measurement of observable in a state", etc. This terminology creates ambiguity because the meaning of the words differs between prevailing quantum mechanics and what is logically and naturally assumed by the human mind in scientific researches and generally used in areas of physics other than quantum mechanics. Nevertheless, I will try using the terminology as close as possible to commonly accepted quantum mechanics paying respects to generations of people who learned quantum mechanics in the existing framework and worked in that area of physics.

Important results like the theorem by Bell [3] and another by Kochen and Specker [4] proved that the existence of non-contextual putative values of observables is in contradiction with the existing formalism of quantum mechanics.

A much more important result is that the existence of non-contextual putative values of observables, when traditionally interpreted, seems to be in contradiction with empirical reality. The common quantum mechanical wisdom reads, for example in the particular case of the position observable, that if we accept that the observable uncertainty is in the system itself, then we must abandon the image of a point and think of it as an extended object: the "particle" is more something like a "field" with all particle properties extended in physical space. This wisdom, as we will see later, must be replaced with more accurate requirement that observables may be placed differently from the states acting on them. Varying of an observable parameters through the physical space will technically be explicitly calculated through the combined states acting on observables.

## 2. Qubit states in geometric algebra

The first step of the long trip to clarity is to follow the definitions strictly:

**Definition 2.1:**

Measurement of observable $O(\cdot)$ in a state $S(\cdot)$ is a map

$$(S(\lambda), O(\mu)) \to O(\nu)$$

where $O(\mu)$ is an element of the set of observables, the values of the elements are identified by some parameters $\mu$, $\nu$, …; $S(\lambda)$ is an element of another set, set of states, the values of the elements are identified by some another parameters $\lambda$, ….

The sets $O(\cdot)$ and $S(\cdot)$ are not necessary different in their formal mathematical implementations. However, alignment in mathematical implementations does not mean that the sets are ontologically identical.

In general, the set of states is external to the set of observables, and vice versa.

**Definition 2.2:**

The result (value) of measurement of observable $O(\cdot)$ in state $S(\cdot)$ is the map sequence

$$(S(\lambda), O(\mu)) \to O(\nu) \to V(B)$$

where $V$ is a set of (Boolean) algebra of subsets identifying possible results of the measurement.

What Dirac had effectively done [5] was to remove the distinction between an element of the operator algebra and the wave function without losing any information about the content of what is carried by the wave function. This is exactly what is shown below to be an accurate implementation of the Definitions 2.1 and 2.2 in the case of a qubit as the state in terms of geometric algebra, when action of a state on observable is non-commutative operation $(S(\lambda), O(\mu)) \to O(\nu) \xleftrightarrow{def} O(\nu) = S^{-1}(\lambda) O(\mu) S(\lambda)$, where $S(\lambda)$ are elements of even subalgebra $G_3^+$ of geometric (Clifford) algebra $G_3$ over three dimensional Euclidean space [2], and $O(\mu)$, $O(\nu)$ are generally elements of $G_3$, though mainly we will consider elements of $G_3^+$.

Another critical thing is explicit generalization of formal "imaginary unit" to a unit value bivector from $G_3^+$ specified by a process under consideration [6] [7]. All that allows to generalize the Dirac's idea and to implement states as the $G_3^+$ valued operators (see also [8]).

To distinguish the $G_3^+$ states from qubits in the $C^2$ Hilbert space I call the former g-qubits [2]. Any $C^2$ qubit $\begin{pmatrix} x_1 + iy_1 \\ x_2 + iy_2 \end{pmatrix}$ has lift in $G_3^+$, namely the g-qubit:

$$x_1 + y_1 B_1 + y_2 B_2 + x_2 B_3 = x_1 + y_1 B_1 + y_2 B_1 B_3 + x_2 B_3 = x_1 + y_1 B_1 + (x_2 + y_2 B_1) B_3 \text{ }^1$$

where $(B_1, B_2, B_3)$ is an arbitrary triple of unit value bivectors in three dimensions satisfying, with the assumption of right-hand screw orientation $B_1 B_2 B_3 = 1^2$, the multiplication rules:

$$B_1 B_2 = -B_3, B_1 B_3 = B_2, B_2 B_3 = -B_1$$

The lift has the $(B_1, B_2, B_3)$ reference frame which can be arbitrary rotated in three dimensions. In that sense we have principal fiber bundle $G_3^+ \to C^2$ with the standard fiber as group of rotations which is also effectively identified by elements of $G_3^+$.

The lift $x_1 + y_1 B_1 + (x_2 + y_2 B_1) B_3$ is the geometric algebra sum of two items, $x_1 + y_1 B_1$ and $(x_2 + y_2 B_1) B_3$, the first is the lift of the quantum mechanical $|0\rangle$, in usual Dirac notations, and the second – lift of $|1\rangle$.

Let the state $x_1 + y_1 B_1$ acts on observable

$$O(\gamma, \delta, I_O) = \gamma + I_O \delta = \gamma + \delta_1 B_1 + \delta_2 B_2 + \delta_3 B_3 \in G_3^+.$$

The item $I_O$ is unit value bivector defining the bivector part of the observable, orientation in three dimension space. If its expansion in the basis is $I_O = \omega_1 B_1 + \omega_2 B_2 + \omega_3 B_3$, $\omega_1^2 + \omega_2^2 + \omega_3^2 = 1$, then $\delta_i = \omega_i \delta, i = 1,2,3$. Thus the action of the state on observable is:

$$(x_1 - y_1 B_1) O(\gamma, \delta, I_O)(x_1 + y_1 B_1)$$

This action does not change the $B_1$ component of the observable and only rotates the remaining of the bivector part belonging to the subspace $span\{B_2, B_3\}$ [2], [7].

---

[1] The lift can equivalently be written as $x_1 + y_1 B_1 + (y_2 - x_2 B_1) B_2$ with identical geometric content.
[2] The reference frame $(B_1, B_2, B_3)$ can be chosen as left-hand screw oriented, $B_1 B_2 B_3 = -1$. It is just reference frame and has nothing to do with physical nature of three dimensional space.

The state $(x_2 + y_2 B_1)B_3$ structurally differs from the lift $x_1 + y_1 B_1$ by additional factor $B_3$. The latter makes flip of the result of the transformation $(x_2 - y_2 B_1)O(\gamma, \delta, I_O)(x_2 + y_2 B_1)$ over the plane $B_1$, particularly changes the sign of the $B_1$ component.

So we get actual geometrical sense of the $G_3^+$ lifts of conventional quantum mechanical basis states $|0\rangle$ and $|1\rangle$. The lift of the first one only rotates observable around the axis orthogonal to basis plane $B_1$, the lift of the second one additionally flips the result, after rotation, over that plane. I will call the states correspondingly as 0-type and 1-type states.

Good to remember that in the geometric algebra formalism measurement of an observable is not distributive relative to linear combinations of states, particularly:

$$(x_1 - y_1 B_1 - B_3(x_2 - y_2 B_1))O(\gamma, \delta, I_O)(x_1 + y_1 B_1 + (x_2 + y_2 B_1)B_3) \neq$$
$$(x_1 - y_1 B_1)O(\gamma, \delta, I_O)(x_1 + y_1 B_1) + (-B_3(x_2 - y_2 B_1))O(\gamma, \delta, I_O)((x_2 + y_2 B_1)B_3)$$

because generally

$$(x_1 - y_1 B_1)O(\gamma, \delta, I_O)((x_2 + y_2 B_1)B_3) + (-B_3(x_2 - y_2 B_1))O(\gamma, \delta, I_O)(x_1 + y_1 B_1) \neq 0$$

Any arbitrary state $s(\alpha, \beta, I_S) = \alpha + \beta_1 B_1 + \beta_2 B_2 + \beta_3 B_3$ can be rewritten either as a 0-type state or 1-type state by rewriting the expression of the bivector part because:

$$\alpha + \beta_1 B_1 + \beta_2 B_2 + \beta_3 B_3 = \alpha + I_{S(\beta_1, \beta_2, \beta_3)} \sqrt{\beta_1^2 + \beta_2^2 + \beta_3^2},$$

where $I_{S(\beta_1, \beta_2, \beta_3)} = \dfrac{\beta_1 B_1 + \beta_2 B_2 + \beta_3 B_3}{\sqrt{\beta_1^2 + \beta_2^2 + \beta_3^2}}$,  0-type,

or

$$\alpha + \beta_1 B_1 + \beta_2 B_2 + \beta_3 B_3 = (\beta_3 + \beta_2 B_1 - \beta_1 B_2 - \alpha B_3)B_3 = \left(\beta_3 + I_{S(\beta_2, -\beta_1, -\alpha)} \sqrt{\alpha^2 + \beta_1^2 + \beta_2^2}\right)B_3$$

where $I_{S(\beta_2, -\beta_1, \alpha)} = \dfrac{\beta_2 B_1 - \beta_1 B_2 - \alpha B_3}{\sqrt{\alpha^2 + \beta_1^2 + \beta_2^2}}$,  1-type.

Choosing of particular flipping plane is not critical since we generally can take the state bivector plane as its only non-zero basis bivector component and write for the corresponding 0-type and 1-type states:

$$\alpha + \beta_1 B_1 + \beta_2 B_2 + \beta_3 B_3 = \alpha + I_{S(\beta_1, \beta_2, \beta_3)} \sqrt{\beta_1^2 + \beta_2^2 + \beta_3^2} =$$
$$\left(\sqrt{\beta_1^2 + \beta_2^2 + \beta_3^2} - I_{S(\beta_1, \beta_2, \beta_3)} \alpha\right) I_{S(\beta_1, \beta_2, \beta_3)}$$

## 3. Geometric algebra states acting at different space points

The transformation $(x_1 - y_1 B_1 - y_2 B_2 - B_3 x_2) O(\gamma, \delta, I_O)(x_1 + y_1 B_1 + y_2 B_2 + B_3 x_2)$ acting on a $G_3^+$ observable $O(\gamma, \delta, I_O)$ assumes the common origin in space for the observable and the state. In other words, the state acting on an observable must be defined at the location of observable. If, for example, the state is rotation of a bivector (observable) represented by a circle, the observable equivalence class member, then the $G_3^+$ state rotates the circle in 3D around its center.

We are targeting to work with anyons, so it is necessary to define the meaning of the $G_3^+$ lifts of quantum mechanical states like $\psi(\vec{r}_1, \vec{r}_2)$ (wave function) or $|\psi_1 \psi_2\rangle$ (Dirac notations) without the assumption of common origin of states and observables.

Suppose we need to transform observable by applying a state, that's mainly to rotate a $G_3^+$ bivector around a point, and the point is different from the observable bivector center. Let the center of rotation is the origin of the coordinate system, null vector, and the observable bivector center is $\vec{r}_0$. All we need is to rotate the observable bivector by given angle around its center, along with the observable center position vector $\vec{r}_0$ rotation by the same angle around the origin. Thus the observable bivector got locally rotated and changed its position in three dimensions.

To make that kind of transformation we need to explicitly write the dependence of the observable on its position is 3D space: $O(\gamma, \delta, I_O) = O(\gamma, \delta, I_O, \vec{r}_0)$. Then the result of transformation is:

$$O^t\left(\gamma, \delta, I_O^t, (x_1 - I_B|\vec{y}|)\vec{r}_0 (x_1 + I_B|\vec{y}|)\right)$$

where $|\vec{y}| = \sqrt{y_1^2 + y_2^2 + y_3^2}$, $I_B = \sum_{i=1}^{3} \frac{y_i}{|\vec{y}|} B_i$, and the explicit transformed value $I_O^t$ can be taken from a bit lengthy formula of the Sec.5.1 of [2]:

$$I_O^t = \delta\left(b_1\left[(x_1^2 + y_1^2) - (y_2^2 + y_3^2)\right] + 2b_2(y_1 y_2 - x_1 y_3) + 2b_3(x_1 y_2 + y_1 y_3)\right)B_1 +$$

$$\delta\left(2b_1(x_1 y_3 + y_1 y_2) + b_2\left[(x_1^2 + y_2^2) - (y_1^2 + y_3^2)\right] + 2b_3(y_2 y_3 - x_1 y_1)\right)B_2 +$$

$$\delta\left(2b_1(y_1 y_3 - x_1 y_2) + 2b_2(x_1 y_1 + y_2 B_3) + b_3\left[(x_1^2 + y_3^2) - (y_1^2 + y_2^2)\right]\right)B_3$$

where $b_i$ are component values of the expansion $I_O = b_1 B_1 + b_2 B_2 + b_3 B_3$ [3]

---

[3] The last formula of $I_O^t$ actually is Hopf fibration received effectively as measurement in geometric algebra terms.

Now assume that we have two observables, $O_1(\gamma_1, \delta_1, I_{O_1}, \vec{r}_1)$ and $O_2(\gamma_2, \delta_2, I_{O_2}, \vec{r}_2)$, and two states $S_1(\alpha^1, \beta^1, I_{S_1})$ and $S_2(\alpha^2, \beta^2, I_{S_2})$. When only one of the observables is in the scene then $S_i(\alpha^i, \beta^i, I_{S_i})$ acts on $O_i(\gamma_i, \delta_i, I_{O_i}, \vec{r}_i)$ with default assumption of coincidence of origins of $S_i(\alpha^i, \beta^i, I_{S_i})$ and $O_i(\gamma_i, \delta_i, I_{O_i}, \vec{r}_i)$. When we have both observables placed at two different space points $\vec{r}_1$ and $\vec{r}_2$, and need to have a compound state assembled of the two states $S_1(\alpha^1, \beta^1, I_{S_1})$ and $S_2(\alpha^2, \beta^2, I_{S_2})$ acting on the observables at that two points we need to also explicitly write the position of a state because the origins of an observable and a state acting on it in measurement may not coincide. Thus we need to explicitly write $S_i(\alpha^i, \beta^i, I_{S_i}) = S_i(\alpha^i, \beta^i, I_{S_i}, \vec{r})$.

The compound state can be easily formalized as:

$$S_{12}(\vec{r}_1, \vec{r}_2) = \int S_1(\alpha^1(\vec{r}), \beta^1(\vec{r}), I_{S_1}(\vec{r})) \delta(\vec{r} - \vec{r}_1) d\vec{r} + \int S_2(\alpha^2(\vec{r}), \beta^2(\vec{r}), I_{S_2}(\vec{r})) \delta(\vec{r} - \vec{r}_2) d\vec{r}$$

where $\delta(\vec{r} - \vec{r}_i)$ is Dirac delta function returning the value of integrated function at point $\vec{r}_i$ from the integral. The functions $\alpha^i(\vec{r})$, $\beta^i(\vec{r})$, $I_{S_i}(\vec{r})$ are arbitrary ones in the three dimensions only satisfying the "pin" requirements $\alpha^i(\vec{r}_i) = \alpha^i$, $\beta^i(\vec{r}_i) = \beta^i$ and $I_{S_i}(\vec{r}_i) = I_{S_i}$.

In a similar way, the system of two observables is:

$$O_{12}(\vec{r}_3, \vec{r}_4) = \int O_1(\gamma_1(\vec{r}), \delta_1(\vec{r}), I_{O_1}(\vec{r}), \vec{r}) \delta(\vec{r} - \vec{r}_3) d\vec{r} + \int O_2(\gamma_2(\vec{r}), \delta_2(\vec{r}), I_{O_2}(\vec{r}), \vec{r}) \delta(\vec{r} - \vec{r}_4) d\vec{r}$$

with the arbitrary functions $\gamma_i(\vec{r})$, $\delta_i(\vec{r})$, $I_{O_i}(\vec{r})$ satisfying $\gamma_i(\vec{r}_i) = \gamma_i$, $\delta_i(\vec{r}_i) = \delta_i$ and $I_{O_i}(\vec{r}_i) = I_{O_i}$.

Assume that $\vec{r}_1 = \vec{r}_3$ and $\vec{r}_2 = \vec{r}_4$, that's the locations of states and of associated observables coincide. Then the system of two observables $O_{12}$ in the state $S_{12}$ returns the following result of measurement:

$$\bar{S}_{12}(\vec{r}_1, \vec{r}_2) O_{12}(\vec{r}_1, \vec{r}_2) S_{12}(\vec{r}_1, \vec{r}_2) =$$
$$\left( \int \bar{S}_1(\alpha^1(\vec{r}), \beta^1(\vec{r}), I_{S_1}(\vec{r})) \delta(\vec{r} - \vec{r}_1) d\vec{r} + \int \bar{S}_2(\alpha^2(\vec{r}), \beta^2(\vec{r}), I_{S_2}(\vec{r})) \delta(\vec{r} - \vec{r}_2) d\vec{r} \right) \times$$
$$\left( \int O_1(\gamma_1(\vec{r}), \delta_1(\vec{r}), I_{O_1}(\vec{r}), \vec{r}) \delta(\vec{r} - \vec{r}_1) d\vec{r} + \int O_2(\gamma_2(\vec{r}), \delta_2(\vec{r}), I_{O_2}(\vec{r}), \vec{r}) \delta(\vec{r} - \vec{r}_2) d\vec{r} \right) \times$$
$$\left( \int S_1(\alpha^1(\vec{r}), \beta^1(\vec{r}), I_{S_1}(\vec{r})) \delta(\vec{r} - \vec{r}_1) d\vec{r} + \int S_2(\alpha^2(\vec{r}), \beta^2(\vec{r}), I_{S_2}(\vec{r})) \delta(\vec{r} - \vec{r}_2) d\vec{r} \right) =$$

$$\bar{S}_1(\alpha^1, \beta^1, I_{S_1}, \vec{r}_1) O_1(\gamma_1, \delta_1, I_{O_1}, \vec{r}_1) S_1(\alpha^1, \beta^1, I_{S_1}, \vec{r}_1) +$$

$$\overline{S}_2(\alpha^2,\beta^2,I_{S_2},\vec{r}_2) O_2(\gamma_2,\delta_2,I_{O_2},\vec{r}_2) S_2(\alpha^2,\beta^2,I_{S_2},\vec{r}_2)+$$

$$\overline{S}_1(\alpha^1,\beta^1,I_{S_1},\vec{r}_1) O_2(\gamma_2,\delta_2,I_{O_2},\vec{r}_2) S_1(\alpha^1,\beta^1,I_{S_1},\vec{r}_1)+$$
$$\overline{S}_2(\alpha^2,\beta^2,I_{S_2},\vec{r}_2) O_1(\gamma_1,\delta_1,I_{O_1},\vec{r}_1) S_2(\alpha^2,\beta^2,I_{S_2},\vec{r}_2)+$$

$$\overline{S}_1(\alpha^1,\beta^1,I_{S_1},\vec{r}_1) O_1(\gamma_1,\delta_1,I_{O_1},\vec{r}_1) S_2(\alpha^2,\beta^2,I_{S_2},\vec{r}_2)+$$
$$\overline{S}_1(\alpha^1,\beta^1,I_{S_1},\vec{r}_1) O_2(\gamma_2,\delta_2,I_{O_2},\vec{r}_2) S_2(\alpha^2,\beta^2,I_{S_2},\vec{r}_2)+$$
$$\overline{S}_2(\alpha^2,\beta^2,I_{S_2},\vec{r}_2) O_1(\gamma_1,\delta_1,I_{O_1},\vec{r}_1) S_1(\alpha^1,\beta^1,I_{S_1},\vec{r}_1)+$$
$$\overline{S}_2(\alpha^2,\beta^2,I_{S_2},\vec{r}_2) O_2(\gamma_2,\delta_2,I_{O_2},\vec{r}_2) S_1(\alpha^1,\beta^1,I_{S_1},\vec{r}_1)$$

The first two members are the results of the action of states on the observables placed at the same positions where the states are initially defined.

The second two members are the results of action of states on the observables located in positions swapped relative to those where the states acting on them are defined. The observables get transformed but also change their positions in three dimensional space. This is explicit demonstration of correctness of the statement from Introduction that a particle (observable) properties are in a sense extendable in physical space.

The last four members are of a sort of transformations different from usual observable measurement transformations and need further detailed elaboration that will be done in a separate research work.

## 4. Geometric algebra anyon exchange statistics in three dimensions

Below is the geometric algebra derivation of the "particle exchange statistics" (see, for example, [9] for the two dimensional case, $|\psi_1\psi_2\rangle = e^{2\pi i\theta}|\psi_2\psi_1\rangle$) in three dimensions. The preceding quotation does not properly reflect physical reality. As we are demonstrating, states are not a particle (observable) internal attributes, rather they are operators acting on observables.

Let's consider the first item of the couple of members from the previous section last formula responsible for applying state action on a varying location observable in physical space:

$$\overline{S}_1(\alpha^1,\beta^1,I_{S_1},\vec{r}_1) O_2(\gamma_2,\delta_2,I_{O_2},\vec{r}_2) S_1(\alpha^1,\beta^1,I_{S_1},\vec{r}_1)$$

Since the state and observable have not coinciding origins we should use the formula given earlier:

$$O^t\left(\gamma, \delta, I_O^t, (x_1 - I_B|\bar{y}|)\vec{r}_0(x_1 + I_B|\bar{y}|)\right)$$

adjusted for the current case. It particularly follows from this formula that the observable location gets moved by the action of the state $S_1\left(\alpha^1, \beta^1, I_{S_1}, \vec{r}_1\right)$ to a new location.

The above formula was derived in assumption that the state is located at the origin of reference system, while the observable is at the position $\vec{r}_0$. In the current case the state $S_1\left(\alpha^1, \beta^1, I_{S_1}, \vec{r}_1\right)$ is located at $\vec{r}_1$ and the observable it acts on is at $\vec{r}_2$. Thus the result becomes:

$$\bar{S}_1\left(\alpha^1, \beta^1, I_{S_1}, \vec{r}_1\right) O_2\left(\gamma_2, \delta_2, I_{O_2}, \vec{r}_2\right) S_1\left(\alpha^1, \beta^1, I_{S_1}, \vec{r}_1\right) =$$

$$O_2^t\left(\gamma_2, \delta_2, I_{O_2}^t, \vec{r}_1 + \bar{S}_1\left(\alpha^1, \beta^1, I_{S_1}, \vec{r}_1\right)(\vec{r}_2 - \vec{r}_1) S_1\left(\alpha^1, \beta^1, I_{S_1}, \vec{r}_1\right)\right)$$

Similarly, for the second term of the considered couple we get:

$$O_1^t\left(\gamma_1, \delta_1, I_{O_1}^t, \vec{r}_2 + \bar{S}_2\left(\alpha^2, \beta^2, I_{S_2}, \vec{r}_2\right)(\vec{r}_1 - \vec{r}_2) S_2\left(\alpha^2, \beta^2, I_{S_2}, \vec{r}_2\right)\right)$$

Suppose now that the two states in the combined state swap their actions on observables:

$$\bar{S}_{12}(\vec{r}_2, \vec{r}_1) O_{12}(\vec{r}_1, \vec{r}_2) S_{12}(\vec{r}_2, \vec{r}_1) =$$
$$\left(\int \bar{S}_1\left(\alpha^1(\vec{r}), \beta^1(\vec{r}), I_{S_1}(\vec{r})\right)\delta(\vec{r} - \vec{r}_2)d\vec{r} + \int \bar{S}_2\left(\alpha^2(\vec{r}), \beta^2(\vec{r}), I_{S_2}(\vec{r})\right)\delta(\vec{r} - \vec{r}_1)d\vec{r}\right) \times$$
$$\left(\int O_1\left(\gamma_1(\vec{r}), \delta_1(\vec{r}), I_{O_1}(\vec{r}), \vec{r}\right)\delta(\vec{r} - \vec{r}_1)d\vec{r} + \int O_2\left(\gamma_2(\vec{r}), \delta_2(\vec{r}), I_{O_2}(\vec{r}), \vec{r}\right)\delta(\vec{r} - \vec{r}_2)d\vec{r}\right) \times$$
$$\left(\int S_1\left(\alpha^1(\vec{r}), \beta^1(\vec{r}), I_{S_1}(\vec{r})\right)\delta(\vec{r} - \vec{r}_2)d\vec{r} + \int S_2\left(\alpha^2(\vec{r}), \beta^2(\vec{r}), I_{S_2}(\vec{r})\right)\delta(\vec{r} - \vec{r}_1)d\vec{r}\right)$$

Then the first couple of terms with not changing observable positions becomes:

$$\bar{S}_1\left(\alpha^1, \beta^1, I_{S_1}, \vec{r}_2\right) O_2\left(\gamma_2, \delta_2, I_{O_2}, \vec{r}_2\right) S_1\left(\alpha^1, \beta^1, I_{S_1}, \vec{r}_2\right) +$$

$$\bar{S}_2\left(\alpha^2, \beta^2, I_{S_2}, \vec{r}_1\right) O_1\left(\gamma_1, \delta_1, I_{O_1}, \vec{r}_1\right) S_2\left(\alpha^2, \beta^2, I_{S_2}, \vec{r}_1\right)$$

Let's consider the difference between $\bar{S}_1\left(\alpha^1, \beta^1, I_{S_1}, \vec{r}_1\right) O_1\left(\gamma_1, \delta_1, I_{O_1}, \vec{r}_1\right) S_1\left(\alpha^1, \beta^1, I_{S_1}, \vec{r}_1\right)$ from the initial case and $\bar{S}_2\left(\alpha^2, \beta^2, I_{S_2}, \vec{r}_1\right) O_1\left(\gamma_1, \delta_1, I_{O_1}, \vec{r}_1\right) S_2\left(\alpha^2, \beta^2, I_{S_2}, \vec{r}_1\right)$ from the last case where the states got swapped.

Rewrite the last item as

$$\bar{S}_2\left(\alpha^2, \beta^2, I_{S_2}, \vec{r}_1\right) S_1\left(\alpha^1, \beta^1, I_{S_1}, \vec{r}_1\right) \times \bar{S}_1\left(\alpha^1, \beta^1, I_{S_1}, \vec{r}_1\right) O_1\left(\gamma_1, \delta_1, I_{O_1}, \vec{r}_1\right) S_1\left(\alpha^1, \beta^1, I_{S_1}, \vec{r}_1\right) \times$$

$$\overline{S}_1(\alpha^1,\beta^1,I_{S_1},\vec{r_1})S_2(\alpha^2,\beta^2,I_{S_2},\vec{r_1})$$

That means that the swapping of states is equivalent to multiplying of $S_1(\alpha^1,\beta^1,I_{S_1},\vec{r_1})$ on the right by $\overline{S}_1(\alpha^1,\beta^1,I_{S_1},\vec{r_1})S_2(\alpha^2,\beta^2,I_{S_2},\vec{r_1})$.

Now comparing the difference between

$\overline{S}_2(\alpha^2,\beta^2,I_{S_2},\vec{r_2})O_2(\gamma_2,\delta_2,I_{O_{S_2}},\vec{r_2})S_2(\alpha^2,\beta^2,I_{S_2},\vec{r_2})$ from the initial case and

$\overline{S}_1(\alpha^1,\beta^1,I_{S_1},\vec{r_2})O_2(\gamma_2,\delta_2,I_{O_2},\vec{r_2})S_1(\alpha^1,\beta^1,I_{S_1},\vec{r_2})$ from the swapped case we similarly get:

$$\overline{S}_1(\alpha^1,\beta^1,I_{S_1},\vec{r_2})S_2(\alpha^2,\beta^1,I_{S_2},\vec{r_2}) \times$$
$$\overline{S}_2(\alpha^2,\beta^2,I_{S_2},\vec{r_2})O_2(\gamma_2,\delta_2,I_{O_2},\vec{r_2})S_2(\alpha^2,\beta^2,I_{S_2},\vec{r_2}) \times \overline{S}_2(\alpha^2,\beta^2,I_{S_2},\vec{r_2})S_1(\alpha^1,\beta^1,I_{S_1},\vec{r_2})$$

which means that the swapping of states is equivalent to multiplying of $S_2(\alpha^2,\beta^2,I_{S_2},\vec{r_2})$ on the right by $\overline{(\overline{S}_1(\alpha^1,\beta^1,I_{S_1},\vec{r_2})S_2(\alpha^2,\beta^2,I_{S_2},\vec{r_2}))}$ which is complex conjugate relative to the result for first two elements.

Similar calculations and comparing the results for the second couples

$$\overline{S}_1(\alpha^1,\beta^1,I_{S_1},\vec{r_1})O_2(\gamma_2,\delta_2,I_{O_2},\vec{r_2})S_1(\alpha^1,\beta^1,I_{S_1},\vec{r_1})+$$
$$\overline{S}_2(\alpha^2,\beta^2,I_{S_2},\vec{r_2})O_1(\gamma_1,\delta_1,I_{O_1},\vec{r_1})S_2(\alpha^2,\beta^2,I_{S_2},\vec{r_2})$$

and

$$\overline{S}_2(\alpha^2,\beta^2,I_{S_2},\vec{r_1})O_2(\gamma_2,\delta_2,I_{O_2},\vec{r_2})S_2(\alpha^2,\beta^2,I_{S_2},\vec{r_1})+$$
$$\overline{S}_1(\alpha^1,\beta^1,I_{S_1},\vec{r_2})O_1(\gamma_1,\delta_1,I_{O_1},\vec{r_1})S_1(\alpha^1,\beta^1,I_{S_1},\vec{r_2})$$

show that the swapping is equivalent to multiplication of $S_1(\alpha^1,\beta^1,I_{S_1},\vec{r_1})$ on the right by $\overline{S}_1(\alpha^1,\beta^1,I_{S_1},\vec{r_1})S_2(\alpha^2,\beta^2,I_{S_2},\vec{r_1})$ and multiplication of $S_2(\alpha^2,\beta^2,I_{S_2},\vec{r_2})$ on the right by $\overline{(\overline{S}_1(\alpha^1,\beta^1,I_{S_1},\vec{r_2})S_2(\alpha^2,\beta^2,I_{S_2},\vec{r_2}))}$. Thus the results are equivalent for both couples of measurement transformations by original and swapped states. So the product $\overline{S}_1(\alpha^1,\beta^1,I_{S_1},\cdot)S_2(\alpha^2,\beta^2,I_{S_2},\cdot) = e^{-I_{S_1}\theta_1}e^{I_{S_2}\theta_2}$, where $\theta_i = \cos^{-1}(\alpha^i)$, is replacing the particle exchange statistics factor $e^{2\pi i \theta}$ when state swapping is generalized to the three dimension anyons in the geometric algebra terms.

Swapping of states acting on observables is one of the two logical options. Another one is swapping locations of two observables. In that case we have:

$$\bar{S}_{12}(\vec{r}_1, \vec{r}_2) O_{12}(\vec{r}_2, \vec{r}_1) S_{12}(\vec{r}_1, \vec{r}_2) =$$
$$\left( \int \bar{S}_1(\alpha^1(\vec{r}), \beta^1(\vec{r}), I_{S_1}(\vec{r})) \delta(\vec{r} - \vec{r}_1) d\vec{r} + \int \bar{S}_2(\alpha^2(\vec{r}), \beta^2(\vec{r}), I_{S_2}(\vec{r})) \delta(\vec{r} - \vec{r}_2) d\vec{r} \right) \times$$
$$\left( \int O_1(\gamma_1(\vec{r}), \delta_1(\vec{r}), I_{O_1}(\vec{r}), \vec{r}) \delta(\vec{r} - \vec{r}_2) d\vec{r} + \int O_2(\gamma_2(\vec{r}), \delta_2(\vec{r}), I_{O_2}(\vec{r}), \vec{r}) \delta(\vec{r} - \vec{r}_1) d\vec{r} \right) \times$$
$$\left( \int S_1(\alpha^1(\vec{r}), \beta^1(\vec{r}), I_{S_1}(\vec{r})) \delta(\vec{r} - \vec{r}_1) d\vec{r} + \int S_2(\alpha^2(\vec{r}), \beta^2(\vec{r}), I_{S_2}(\vec{r})) \delta(\vec{r} - \vec{r}_2) d\vec{r} \right) =$$

$$\bar{S}_1(\alpha^1, \beta^1, I_{S_1}, \vec{r}_1) O_2(\gamma_2, \delta_2, I_{O_2}, \vec{r}_1) S_1(\alpha^1, \beta^1, I_{S_1}, \vec{r}_1) +$$
$$\bar{S}_2(\alpha^2, \beta^2, I_{S_2}, \vec{r}_2) O_1(\gamma_1, \delta_1, I_{O_1}, \vec{r}_2) S_2(\alpha^2, \beta^2, I_{S_2}, \vec{r}_2) +$$

$$\bar{S}_1(\alpha^1, \beta^1, I_{S_1}, \vec{r}_1) O_1(\gamma_1, \delta_1, I_{O_1}, \vec{r}_2) S_1(\alpha^1, \beta^1, I_{S_1}, \vec{r}_1) +$$
$$\bar{S}_2(\alpha^2, \beta^2, I_{S_2}, \vec{r}_2) O_2(\gamma_2, \delta_2, I_{O_2}, \vec{r}_1) S_2(\alpha^2, \beta^2, I_{S_2}, \vec{r}_2) +$$

$$\bar{S}_1(\alpha^1, \beta^1, I_{S_1}, \vec{r}_1) O_1(\gamma_1, \delta_1, I_{O_1}, \vec{r}_2) S_2(\alpha^2, \beta^2, I_{S_2}, \vec{r}_2) +$$
$$\bar{S}_1(\alpha^1, \beta^1, I_{S_1}, \vec{r}_1) O_2(\gamma_2, \delta_2, I_{O_2}, \vec{r}_1) S_2(\alpha^2, \beta^2, I_{S_2}, \vec{r}_2) +$$
$$\bar{S}_2(\alpha^2, \beta^2, I_{S_2}, \vec{r}_2) O_1(\gamma_1, \delta_1, I_{O_1}, \vec{r}_2) S_1(\alpha^1, \beta^1, I_{S_1}, \vec{r}_1) +$$
$$\bar{S}_2(\alpha^2, \beta^2, I_{S_2}, \vec{r}_2) O_2(\gamma_2, \delta_2, I_{O_2}, \vec{r}_1) S_1(\alpha^1, \beta^1, I_{S_1}, \vec{r}_1)$$

Calculations similar to the case of swapping states give the same result $e^{-I_{S_1}\theta_1} e^{I_{S_2}\theta_2}$ for the generalized three dimensional anyons exchange statistics factor.

## 5. Conclusions

Generalization of the $C^2$ Hilbert space qubit formalism to the even subalgebra $G_3^+$ in three dimensions formalism [2]

- opens new ways to deal with anyons;

- and subsequently supports more profound braiding formalism in three dimensions;
- and in that way is expanding the horizon of topological quantum computing implementation.